\documentclass[a4paper,12pt]{article}
\usepackage[utf8]{inputenc}
\usepackage[english]{babel}
\usepackage{graphicx}
\usepackage{subcaption}
\usepackage{mathtext}
\usepackage{braket}
\usepackage{feynmp-auto}
\usepackage{indentfirst}
\usepackage{epsfig,amsmath,amsfonts,braket,fp,bbm,tikz-cd,authblk,hyperref}
\usepackage{amssymb}
\usepackage{amsmath}
\usepackage{slashed}
\usepackage{euscript}
\usepackage{bbold}
\usepackage{bm,paralist,xspace,url,relsize}
\usepackage{color,xcolor}
\usepackage{hyperref}
\usepackage{textcomp}
\usepackage{appendix}
\usepackage{titlesec}
\usepackage{blindtext}

\usepackage{pgfplots}
\pgfplotsset{compat=1.13}
\usetikzlibrary{calc}
\usepackage{tikz}

\usepackage[font=normalsize,labelfont=,labelsep=colon]{caption}
\definecolor{mypink1}{rgb}{0.858, 0.188, 0.478}

\usepackage{cmap} 
\usepackage[left=1.5cm,right=1.5cm,top=2cm,bottom=2cm,bindingoffset=0cm]{geometry}
\PassOptionsToPackage{unicode}{hyperref}
\PassOptionsToPackage{naturalnames}{hyperref}
\usepackage{titlesec}
\titleformat{\section}{\normalfont\large\bfseries}{\thesection}{1em}{}
\titleformat{\subsection}{\normalfont\normalsize\bfseries}{\thesubsection}{1em}{}

\usepackage[active]{srcltx}

\usepackage{tikz}

\begin{document}
\setcounter{section}{0}
\renewcommand{\contentsname}{Contents}

\author[]{ O.Diatlyk}
\affil[]{\small National Research University Higher School of Economics\\ Myasnitskaya Ulitsa, 20, 101000 Moscow, Russia}

\title{\huge Hawking radiation of massive fields in 2D}
\date{}

\numberwithin{equation}{section}

\maketitle
\vspace{3cm}
\begin{abstract}
We extend the classic results of the paper P. C. W. Davies, S. A. Fulling, and W. G. Unruh "Energy-momentum tensor near an evaporating black hole" by considering a massive scalar field in a two dimensions in the presence of a thin shell collapse. We show that outside the shell the WKB approximation is valid for any value of $r$ if $mr_{g} \gg 1$, where $m$ is the mass of the field, and $r_{g}$ is the Schwarzschild radius. Thus, we use semiclassical modes to calculate the flux in the vicinity of the shell, and at spatial infinity, $r \rightarrow +\infty$ at the final stage of the collapse, $t \rightarrow +\infty$ with the use of the covariant point-splitting regularization. We get that near the shell and at the spatial infinity the radiation is thermal with Hawking temperature. We obtain the negative flux $T_{vv}$ in the vicinity of the shell, which is similar to the classic result in the massless case. 
\end{abstract}
\newpage
\tableofcontents

\newpage

\section{Introduction}

In this paper we extend the classic results of P. C. W. Davies, S. A. Fulling, and W. G. Unruh \cite{Davies:1976ei} to the case of the massive scalar field theory on the background of a collapsing thin shell in two dimensions. The background is described in details in \cite{Unruh:1976db},\cite{Akhmedov:2015xwa} but we briefly review the main points in the section \ref{TBGaWA} for completeness of the paper. Our goal is to calculate the expectation value of the stress-energy tensor at the final stage of the thin shell collapse using the covariant point-splitting  regularization \cite{Davies covariant point splitting}-\cite{Akhmedov:2020ryq}.

In this classic paper \cite{Davies:1976ei}, the authors calculate the expectation value, $T_{\mu\nu}$, of the stress-energy tensor for the massless scalar field in the two-dimensional model of the gravitational collapse. In two dimensions, there is a problem that  Hawking radiation is incompatible with the conserved and traceless $T_{\mu\nu}$. Depending on the scheme of regularization they got either conserved or traceless stress energy tensor. 
From the physical point of view we need to demand the conservation of stress energy tensor in order to generalize these results to four dimensions. It supports the picture of the black hole evaporation in which pairs of particles are created outside the horizon (and not entirely in the collapsing matter), one of which carries negative energy towards the future horizon of the black hole, while the other contributes to the thermal flux at infinity:
\begin{eqnarray}
     \label{eq:58}
     \left\{
     \begin{matrix}
      T_{vv}=-\dfrac{1}{192 \pi r_{g}^{2}}, &  T_{uu}=0,   & \text{as }\qquad r\rightarrow{r_{g}}, \\
    T_{vv}=0, & T_{uu}=\dfrac{1}{192 \pi r_{g}^{2}}, & \text{as} \qquad r\rightarrow{\infty}.
     \end{matrix}\right.
 \end{eqnarray} 
Such $T_{\mu\nu}$ will lead to the evaporation of the horizon during the Hawking radiation during the thin shell collapse. However, massive theory in 2D is not conformally invariant and the method of \cite{Davies:1976ei} could not be used in this case. The goal of this paper is to fill in this gap.

We calculate the expectation value of the stress energy tensor in the same  background, discussed above, for the massive case. Unfortunately, the model is not solvable exactly. However, we show that in the limit of the heavy fields, $mr_{g} \gg 1$, where $m$ is the mass of the scalar field, and $r_{g}$ is the Schwarzschild radius, we can find modes outside the shell for all values of $r$ with the use of the WKB method. 
Then, we calculate the expectation value of the stress energy tensor $T_{\mu\nu}$ using the same covariant point-splitting regularization and WKB modes in the vicinity of the shell and at spatial infinity as $t \rightarrow +\infty$. 

The paper is organized as follows. In the section \ref{TBGaWA} we briefly discuss the geometry of the thin shell collapse and the behaviour of the massive scalar field in such a background. In the limit $mr_{g} \gg 1$, we can find modes outside the shell for any value of $r$ using the WKB method.

In section \ref{Modesbefore} we find in-harmonics -- modes
that diagonalize the free hamiltonian before the start of the collapse, when the shell is stationary, hence providing a sensible definition of the in ground state. The state with respect to which the flux is found is defined in terms of these modes. In section \ref{Modesafter} we find the behaviour of in-harmonics at the final stage of the thin
shell collapse (for more details, see \cite{Akhmedov:2015xwa}).

In  section \ref{SETafter} we obtain covariantly conserved stress energy tensor for the massive theory at the final stage of the collapse at the spatial infinity ($t,r_{*} \rightarrow \infty$, such that $|t-r_{*}| < \infty$), using the covariant point-splitting regularization. We obtain thermal behaviour for the $T_{tr_{*}}$ component. 
In section \ref{SETbefore} we discuss the results of the calculation of covariantly conserved stress energy tensor for a massive scalar field in the vicinity of the shell and compare it with results of the paper \cite{Davies:1976ei}.  Near the shell we obtain the same thermal radiation as at spatial infinity. Even though we get the same thermal flux $T_{tr_{*}}=T_{uu}-T_{vv}$ near the horizon and at spatial infinity, still the components $T_{uu}$ and $T_{vv}$ are different: at infinity we find out going flux, while near the horizon there is a negative ingoing flux.

We discuss the results and the future steps in the section \ref{Conclusions}. To make the paper self-contained  the details of calculations are present in the Appendix \ref{Calculations}-\ref{SEThorizon}.

\section{The background geometry and WKB approach}\label{TBGaWA}
The two-dimensional metric, which we use as the background for the massive scalar field theory has the following form 
\begin{equation}
\label{eq:1}
ds^2=
\left\{\begin{matrix}
dt^{2}_{-}-dr^{2}, & r<R(t), \\
\left(1-\dfrac{r_{g}}{r}\right)dt^{2}-\dfrac{dr^{2}}{1-\dfrac{r_{g}}{r}}, &  r >R(t).
\end{matrix}\right.
\end{equation}
Before the collapse, $t \leq0$, the shell is at rest: $R(t)=R_{0}$ and $0<R_{0}-r_{g} \ll r_{g}$. In this case, the times inside and outside the shell are related as:
\begin{equation}
    \label{eq:2}
    t_{-}=t\sqrt{1-\dfrac{r_{g}}{R_{0}}},  \qquad t<0.
\end{equation}
At the final stage of the collapse, which starts at $t=0$, the shell’s trajectory from the point of view of the outside observer is:
\begin{equation}
\label{eq:8}
R(t) \approx r_{g}\left(1+\frac{R_{0}-r_{g}}{r_{g}}e^{-\frac{t}{r_{g}}} \right).
\end{equation}
In terms of tortoise coordinates, $r_{*}=r+r_{g}\log \left(\frac{r}{r_{g}}-1\right)$, the trajectory has the form:
\begin{equation}
\label{eq:9}
R_{*}(t)\approx R_{0*}-t+\left(r_{g}-R_{0}\right)\left(1-e^{-\frac{t}{r_{g}}}\right).
\end{equation}
For the internal observer, given the assumption that $\left|R(t)-r_{g}\right| \ll r_{g}$ the shell collapses at almost constant speed $c$, which is defined as
$$c \approx \left|\frac{dR(t_{-})}{dt_{-}}\right|\approx 1,\quad R\left(t_{-}\right) \approx R_{0}-ct_{-}.$$
Then the relation between the times outside and inside the shell during this stage of the collapse process is
\begin{equation}
\label{eq:10}
t_{-}\approx\frac{R_{0}-r_{g}}{c}\left(1-e^{-\frac{t}{r_{g}}}\right), \quad t\to\infty.
\end{equation}
For the details and some further discussions see \cite{Unruh:1976db},\cite{Akhmedov:2015xwa}. 

The theory that we consider on this background is as follows:
\begin{equation}
    \label{EQ:o1}
    S=\dfrac{1}{2}\int d^{2}x\sqrt{|g|}\left[(\partial_{\mu}\phi)^{2}-m^{2}\phi^{2}\right].
\end{equation}
Using the exact form of the background metric \eqref{eq:1}, the action takes the form
\begin{gather}
\label{eq:11}
  S=\dfrac{1}{2} \int dt \int \limits_{0}^{R(t)} dr \dfrac{\partial t_{-}}{\partial t}\left[\left(\frac{\partial t}{\partial t_{-}}\right)^2(\partial_{t}\phi)^2-(\partial_{r}\phi)^2-m^{2}\phi^{2}\right]+ \hspace{5cm}\notag\\ \hspace{5cm}+\dfrac{1}{2}\int dt \int \limits_{R(t)}^{+\infty}dr\left[\frac{(\partial_{t}\phi)^2}{1-\dfrac{r_{g}}{r}}-\left(1-\frac{r_{g}}{r}\right)(\partial_{r}\phi)^2-m^{2}\phi^{2}\right],
\end{gather}
Varying this action we obtain the equations of motions
\begin{gather}
    \label{EQ:o2}
   \left\{ \begin{matrix}
\left[\partial^{2}_{t_{-}}-\partial^{2}_{r}-m^{2}\right]\phi=0 , & r<R(t),\\
  \left[ \partial^{2}_{t}-\partial^{2}_{r_{*}}+\left(1-\dfrac{r_{g}}{r}\right)m^{2}\right]\phi=0, & r>R(t),
    \end{matrix}\right.
\end{gather}
with gluing conditions  at the shell as follows\footnote{for simplicity we use that $\phi(r=0)=0$, even though it is possible to generalize our calculation to different boundary conditions \cite{Unruh:1976db}}:
\begin{gather}
\phi(r=0)=0,\quad \phi \left(R(t)-\epsilon \right)=\phi \left(R(t)+\epsilon \right),\quad \text{and} \notag\\
     \left[\left(\frac{\partial t}{\partial t_{-}}\right)\left|\frac{dR(t)}{dt}\right|\partial_{t}\phi-\dfrac{\partial t_{-}}{\partial t}\partial_{r}\phi\right]_{R(t)-\epsilon}=\left[\frac{\partial_{t}\phi}{1-\frac{r_{g}}{r}}\left|\frac{dR(t)}{dt}\right|-\left(1-\dfrac{r_{g}}{r}\right)\partial_{r}\phi\right]_{R(t)+\epsilon}. \label{eq:13b}
 \end{gather}

Inside the shell the harmonics  are plane waves as follows from  \eqref{EQ:o2}. Let us analyze the equations of motion outside the shell. First of all, the term $1-\dfrac{r_{g}}{r(r_{*})}$ can be represented via the Lambert function as
\begin{equation}
    \label{eq:145}
    1-\dfrac{r_{g}}{r(r_{*})}=\dfrac{W \left(e^{r_{*}/r_{g}-1}\right)}{1+W \left(e^{r_{*}/r_{g}-1}\right)}.
\end{equation}
Before the collapse we can represent $\phi(r,t)=e^{-i\omega t}\phi_{\omega}(r)$, and obtain the following equation for $\phi_{\omega}(r)$:
\begin{eqnarray}
    \label{eq:146}
    \left[\partial^{2}_{r_{*}}+\omega^{2}-V[r_{*}]\right]\phi_{\omega}=0, \qquad \qquad V[r_{*}]=m^{2}\left(1-\dfrac{r_{g}}{r(r_{*})}\right)
\end{eqnarray}
or in terms of dimensionless argument $x\equiv\dfrac{r_{*}}{r_{g}}$:
\begin{eqnarray}
    \label{eq:147}
    \left[\lambda^{2}\partial^{2}_{x}+q^{2}(x)\right]\phi_{\omega}=0,
\end{eqnarray}
where 
\begin{gather}
    \label{eq:148}
    \left\{ \begin{matrix}
    \lambda=\dfrac{1}{mr_{g}},\\
    q^{2}(x)=\dfrac{\omega^{2}}{m^{2}}-\left(1-\dfrac{r_{g}}{r[r_{g}x]}\right).
    \end{matrix}\right.
\end{gather}


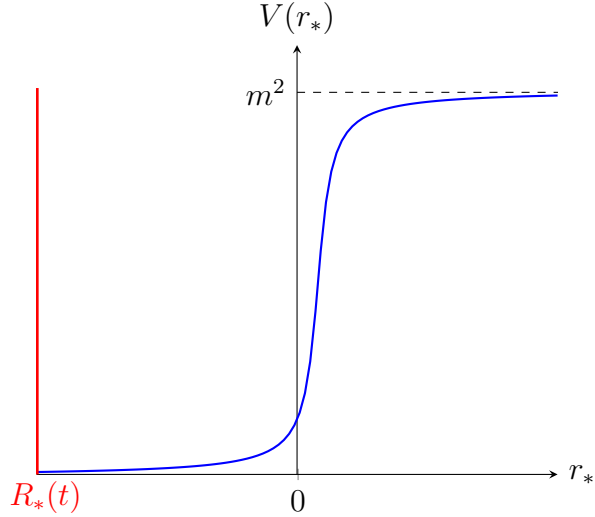
\begin{figure}
\centering
\begin{tikzpicture}
\begin{axis}[
	axis x line=center,
	axis y line=center,
	domain=-25:25,
	xlabel=$r_{*}$, xlabel style={at=(current axis.right of origin), anchor= west},
	ylabel=$V(r_{*})$, ylabel style={at=(current axis.above origin), anchor=south },
	samples=100,
	ymin=-0.02,ymax=200,xmin=-25,xmax=25, xtick={0.1}, xticklabels={$0$}, ytick=\empty ]
	  \addplot+[mark=none, draw=blue, thick] {atan(x-2) + 89};

    \draw [black, dashed, mark size=3 pt] (axis cs:+50,+178) -- (axis cs: 0,+178) node [left]{$m^{2}$};
    \draw [red, mark size=3 pt, ultra thick] (axis cs:-25,+0)  -- (axis cs: -25,+180) node [above right]{$$};
\end{axis}
\node [color=red, right]at (-.5,-.3) {$R_{*}(t)$};
\end{tikzpicture}
\caption{The form of the potential $V[r_{*}]$ in \eqref{eq:146}. The limit $r_{*} \rightarrow -\infty$ corresponds to $r \rightarrow r_{g}$. The red line corresponds to the position of the shell. At the final stage of the collapse process the shell surface is moving towards infinity: $R_{*}(t) \rightarrow -\infty$. }
\label{fig10dm}
\end{figure}
Because of the form of the potential $V(r_{*})$ it is convenient to seperate solutions of the equation \eqref{eq:147} in two regions: the modes with $m\sqrt{1-\frac{r_{g}}{R_{0}}}<\omega<m$, which oscillate as $r_{*}\rightarrow -\infty$ and exponentially decay as $r_{*} \rightarrow +\infty$, and the modes with $m<\omega<+\infty$, which oscillate for any value of $r_{*}$ (see fig. \ref{fig10dm}).
If one assumes that $\lambda \ll 1$, then the condition of the validity of the WKB approach, i.e.
\begin{equation}
    \label{eq:149}
    \dfrac{1}{2mr_{g}}\dfrac{1}{\left[\omega^{2}/m^{2}-\left(1-r_{g}/r\right)\right]^{3/2}}\dfrac{r^{2}_{g}}{r^{2}}\left[1-\dfrac{r_{g}}{r}\right] \ll 1,
\end{equation}
is fulfilled for any $r$ and $\omega>m$. We discuss the harmonics with $m\sqrt{1-\frac{r_{g}}{R_{0}}}<\omega<m$ below.

The above-mentioned statement means that any mode can be represented as 
\begin{gather}
   \phi_{\omega>m}(t,r_{*})=   \label{eq:150}\\
   =\int \limits_{|\omega^{\prime}|>m}\dfrac{d\omega^{\prime}}{2\pi}\bigg(C_{1}(\omega,\omega^{\prime})\dfrac{1}{\sqrt{k^{\prime}(r_{*})}}e^{-i\omega^{\prime}t-i\int \limits_{R_{0*}}^{r_{*}}dxk^{\prime}(x)}+C_{2}(\omega,\omega^{\prime})\dfrac{1}{\sqrt{k^{\prime}(r_{*})}}e^{-i\omega^{\prime} t+i\int \limits_{R_{0*}}^{r_{*}}dxk^{\prime}(x)}\bigg),\ \  r>R(t),
  \notag
\end{gather}
where 
\begin{eqnarray}
    \label{eq:151}
    k^{2}(r_{*})=\omega^{2}-m^{2}\left(1-\dfrac{r_{g}}{r(r_{*})}\right).
\end{eqnarray}
It is worth to mention that the integration in \eqref{eq:150} is over both positive and negative $\omega^{\prime}$. The form of \eqref{eq:150} means that in general we have some linear combination of the exponents of the form 
\begin{equation*}
\exp\left(\pm i\omega' t \pm i\int \limits_{R_{0*}}^{r_{*}}dxk^{\prime}(x)\right).
\end{equation*}
For example, before the start of the collapse the situation is stationary,  $C_{1}(\omega,\omega^{\prime}) \sim \delta(\omega-\omega^{\prime})$, $C_{2}(\omega,\omega^{\prime}) \sim \delta(\omega-\omega^{\prime})$. At the final stage of the collapse, as we show later, such a separation of $t$ and $r$ is not possible because the background depends on time, and the harmonics have a more complicated form.
Using these semiclassical harmonics in the next sections we find the expectation value of the stress energy tensor near the shell and at spatial infinity, as $t \rightarrow +\infty$.

Thus, we get that the harmonics are
\begin{gather}
    \phi_{\omega>m}= 
   \left\{ 
   \begin{matrix}
    \dfrac{1}{\sqrt{2\omega}}e^{-i\omega_{-}t_{-}}\left(e^{-ik_{-}r}-e^{ik_{-}r}\right),  \ \  r<R(t),  \\ 
 {\displaystyle \int \limits_{|\omega'|>m} }\dfrac{d\omega^{\prime}}{2\pi}\left(\dfrac{C_{1}(\omega,\omega^{\prime})}{\sqrt{k^{\prime}(r_{*})}}e^{-i\omega^{\prime}t-i\int \limits_{R_{0*}}^{r_{*}}dxk^{\prime}(x)}+\dfrac{C_{2}(\omega,\omega^{\prime})}{\sqrt{k^{\prime}(r_{*})}}e^{-i\omega^{\prime} t+i\int \limits_{R_{0*}}^{r_{*}}dxk^{\prime}(x)}\right),\ \  r>R(t).
    \end{matrix}\right.
\end{gather}
Using gluing conditions \eqref{eq:13b} we can find unknown coefficients $C_{1}(\omega,\omega^{\prime})$ and $C_{2}(\omega,\omega^{\prime})$.\


\section{Modes}
The set up of the problem is as follows. We solve Klein Gordon equation \eqref{EQ:o2},  with the boundary conditions \eqref{eq:13b} for any $r$ and any $t$. The solution of this equation is some complicated function $\phi(r,t)$ for which we cannot separate the dependence on $r$ and $t$ as $\phi_{\omega}(r)e^{i\omega t}$ for all times.  This is a very complicated problem.  

However, one can find an approximate solution of this problem in two different phases.
First phase, is defined for $t<0$ -- the shell is at rest. We use the in-harmonics -- modes which diagonilize the free hamiltonian. In this phase the modes have the form $\phi_{\omega}(r)e^{i\omega t}$. The phase II is the late stage of collapse when $R(t) \rightarrow r_{g}$ as $t \rightarrow +\infty$. At this phase we look for the approximate solution of the in-harmonics at $t\rightarrow+\infty$. The form $\phi_{\omega}(r)e^{i\omega t}$, $t \leq 0$ is used as initial value for the second stage.

 \subsection{The behaviour of in-harmonics before the collapse}\label{Modesbefore}
In this section we find harmonics before the collapse. During the first stage, when the shell is stationary, we can find in-harmonics
that diagonalize the free hamiltonian.
As was mentioned in the section \ref{TBGaWA}, before collapse the time inside and outside the shell are related according to \eqref{eq:2},
which means that the modes inside the shell have the form\footnote{For $\omega >m$ the coefficient $\left(1-r_{g}/R_{0}\right)^{-1/4}\dfrac{1}{\sqrt{2k_{-}}}$ reduces to $\dfrac{1}{\sqrt{2\omega}}$ since $R_{0}-r_{g} \ll r_{g}$.}
\begin{eqnarray}
    \label{que:8}
    \phi_{\omega}(t,r)=\left(1-r_{g}/R_{0}\right)^{-1/4}\dfrac{1}{\sqrt{2k_{-}}}e^{-i\omega_{-}t_{-}}\left(e^{-ik_{-}r}-e^{ik_{-}r}\right),  \ \  r<R_{0},
\end{eqnarray}
with $\omega_{-}=\omega/\sqrt{1-r_{g}/R_{0}}$ and $k_{-}=\sqrt{\omega^{2}_{-}-m^{2}}$. From \eqref{que:8} we can see that harmonics are bounded from below: $\omega>m\sqrt{1-\frac{r_{g}}{R_{0}}}$. 

As is usual in canonical quantization, we expand the scalar field in terms of a basis of harmonics 
\begin{equation}
    \label{EQrrr:20}
    \hat{\phi}=\int \limits_{m\sqrt{1-r_{g}/R_{0}}}^{+\infty} \dfrac{d\omega}{2\pi}\big(\phi_{\omega}\hat{a}_{\omega}+\text{h.c.}\big), \qquad [\hat{a}_{\omega^{\prime}},\hat{a}_{\omega^{\prime\prime}}^{\dagger}]=2\pi\delta(\omega^{\prime}-\omega^{\prime\prime}),
\end{equation}
where $\omega$ is the quantum number
that labels the harmonics, and plays the role of energy before the collapse. 
The harmonics inside the shell satisfy the following commutation relation:
\begin{equation}
    \label{eq:119}
     g^{tt}\int \limits_{m\sqrt{1-r_{g}/R_{0}}}^{+\infty}\dfrac{d\omega}{2\pi}\left(\phi_{\omega}(t,r)\partial_{t}\phi_{\omega}^{*}(t,r^{\prime})- \text{h.c.}\right)=i\delta \left(r-r^{\prime}\right)-i\delta \left(r+r^{\prime}\right),
\end{equation}
where in addition to the usual delta-function appears the boundary one $\delta(r+r^{\prime})$ whose argument is equal to
zero only at the boundary $r=r^{\prime}=0$. Since before the collapse everything is stationary we look for the solution outside the shell in the form 
\begin{eqnarray}
    \label{que:9}
    \phi_{\omega}(t,r_{*})=\dfrac{C_{1}}{\sqrt{2k(r_{*})}}e^{-i\omega t-i\int \limits_{R_{0*}}^{r_{*}}dxk(x)}+\dfrac{C_{2}}{\sqrt{2k(r_{*})}}e^{-i\omega t+i\int \limits_{R_{0*}}^{r_{*}}dxk(x)}.
\end{eqnarray} 
To find coefficients $C_{1}$ and $C_{2}$ one needs to use the gluing conditions at $R_{0}$ \eqref{eq:13b}. Using that $k^{2}(r_{*}) \rightarrow \omega^{2}$ when $r \rightarrow r_{g}$, we have that \begin{gather}
    \label{que:10}
  \left\{  \begin{matrix}
    C_{1}=e^{-ik_{-}R_{0}}, \\
    C_{2}=-e^{ik_{-}R_{0}}.
    \end{matrix}\right.
\end{gather}
Finally, 
\begin{gather}
    \label{que:11}
    \phi_{\omega>m}(t,r)=
\left \{
\begin{matrix}
    \dfrac{1}{\sqrt{2\omega}}e^{-i\omega_{-}t_{-}}\left(e^{-ik_{-}r}-e^{ik_{-}r}\right),  &  r<R_{0}, \\
    e^{-ik_{-}R_{0}}\dfrac{1}{\sqrt{2k(r_{*})}}e^{-i\omega t-i\int \limits_{R_{0*}}^{r_{*}}dxk(x)}-e^{ik_{-}R_{0}}\dfrac{1}{\sqrt{2k(r_{*})}}e^{-i\omega t+i\int \limits_{R_{0*}}^{r_{*}}dxk(x)}, & r>R_{0},
    \end{matrix}\right.
\end{gather}
For future reference, it is worth to mention that the harmonics outside the shell with $r-R_{0} \ll r_{g}$ have the form 
\begin{eqnarray}
    \label{que:12}
    \phi_{\omega } \approx \dfrac{1}{\sqrt{2\omega}}e^{-i\omega v+i\varphi_{\omega }/2}-\dfrac{1}{\sqrt{2\omega}}e^{-i\omega u-i\varphi_{\omega }/2},
\end{eqnarray}
with 
\begin{gather}
    \label{que:13}
   \left\{ \begin{matrix}
    v=t+r_{*},\\
    u=t-r_{*}, \\
    \varphi_{\omega }=2\left(\omega R_{0*}-k_{-}R_{0}\right)\approx 2\omega \left(R_{0*}-R^{-}_{0}\right),
    \end{matrix}\right. \qquad \text{and} \quad R^{-}_{0}=\dfrac{R_{0}}{\sqrt{1-r_{g}/R_{0}}} \rightarrow \infty.
\end{gather}
 We will need these expressions when in the next section will look for the in-harmonics outside the shell at the final stage of the collapse, i.e. as $t \rightarrow +\infty$. 
 
 For completeness, we mention that in the vicinity of the shell with the radius $R_{0}$ there is a discrete spectrum with $m\sqrt{1-r_{g}/R_{0}}<\omega<m$,  \cite{Akhmedov:2016uha}. Harmonics from this part of the spectrum exponentially decay at spatial infinity. We do not provide the exact form of those harmonics because we do not discuss the expectation value of the stress energy tensor before the collapse. However, at the final stage of the collapse the harmonics with $\omega<m$ play an important role near the shell and we will discuss them below.

\subsection{In-harmonics during the late-stage of the collapse}\label{Modesafter}

The calculation of the in-harmonics as $t \rightarrow +\infty$  in the vicinity of the shell is similar to the one in \cite{Akhmedov:2015xwa}. Hence, we state here only the result:
\begin{gather}
\label{eq:139}
    \phi_{\omega>m} \approx
   \left\{ \begin{matrix}
    \dfrac{e^{-i\omega _{-}t_{-}}}{\sqrt{2\omega}}\left(e^{-ik_{-}r}-e^{ik_{-}r}\right), & r<R(t), \\
    \frac{1}{\sqrt{2\omega}}e^{-i\omega v+i\varphi_{\omega}/2}-\frac{2i}{\sqrt{2\omega}}e^{-i\omega (BU+A)}\sin{\omega_{-}r_{g}},  &  r>R(t), \ |R(t)-r_{g}| \ll r_{g},
    \end{matrix}\right.
\end{gather}
with
\begin{gather}
    \label{eq:178}
  \left\{  \begin{matrix}
    U=e^{-u/2r_{g}}, \\
   \varphi_{\omega}= 2\omega\left(R^{*}_{0}-R^{-}_{0}\right),\\
B=-\frac{(R_{0}-r_{g})e^{-\frac{R^{*}_{0}+r_{g}-R_{0}}{2}}}{\sqrt{1-\frac{r_{g}}{R_{0}}}} \approx -r_{g}e^{-1/2},\\ A=\frac{R_{0}-r_{g}}{\sqrt{1-\frac{r_{g}}{R_{0}}}} \rightarrow 0,
    \end{matrix}\right.
\end{gather}
if $c \approx 1$, i.e. the shell is almost light-like at the late-stage of collapse, the $v$-dependent part of the modes outside the
shell for $r\rightarrow r_{g}$ is not affected by the collapse (see equation \eqref{que:12}). For convenience, we can reexpand the $u$-dependent part as
\begin{eqnarray}
    \label{eq:179}
    -\dfrac{2i}{\sqrt{2\omega}}e^{-i\omega(BU+A)}\sin{\omega_{-}r_{g}}=\int \limits_{|\omega^{\prime}|>m}\dfrac{d\omega^{\prime}}{2\pi}\dfrac{1}{\sqrt{2\omega^{\prime}}}\gamma(\omega,\omega^{\prime})e^{-i\omega^{\prime}u},
\end{eqnarray}
where
\begin{eqnarray}
    \label{eq:180}
    \gamma(\omega,\omega^{\prime}) \approx -4ir_{g}\sqrt{\dfrac{|\omega^{\prime}|}{\omega}}\sin{\left(\omega_{-}r_{g}\right)}e^{-i\omega A}e^{-ir_{g}\omega^{\prime}}e^{\pi r_{g}\omega^{\prime}}e^{2ir_{g}\omega^{\prime}\log{\omega r_{g}}} \Gamma\left(-2i\omega^{\prime}r_{g}\right).
\end{eqnarray}

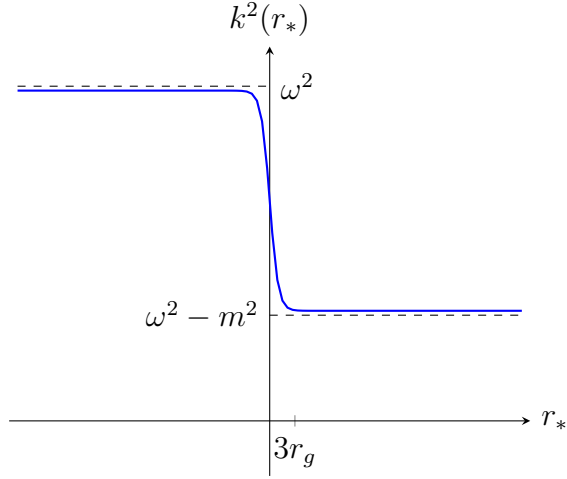
\begin{figure}[t!]
\centering
\begin{tikzpicture}
\begin{axis}[
	axis x line=center,
	axis y line=center,
	domain=-30:30,
	xlabel=$r_{*}$, xlabel style={at=(current axis.right of origin), anchor= west},
	ylabel=$k^{2}(r_{*})$, ylabel style={at=(current axis.above origin), anchor=south },
	samples=100,
	ymin=-0.5,ymax=3.4,xmin=-31,xmax=31,
	xtick={0,3},xticklabels={0,$3r_{g}$},ytick=\empty]
	  \addplot+[mark=none, draw=blue, thick] {-tanh(\x)+2 };
    

    \draw [black, dashed, mark size=3 pt] (axis cs:0,+0.96) node [left]{$\omega^{2}-m^{2}$}-- (axis cs:+30,+0.96);
    \draw [black, dashed, mark size=3 pt] (axis cs:-30,+3.04) -- (axis cs: 0,+3.04) node [right]{$\omega^{2}$};
\end{axis}
\end{tikzpicture}
\caption{This is a graph of $k^{2}(r_{*})$ as function of $r_{*}$ coordinate. As could be seen, for $r_{*} \gg 1$ the function $k^{2}(r_{*})$ can be approximated as a constant $\omega^{2}-m^{2}$. On the other hand, as $r_{*}\rightarrow -\infty$ (which is the same as $r \rightarrow r_{g}$) the function $k^{2}(r_{*})$ takes its maximum value $\omega^{2}$. Finally, we can approximate $k^{2}(r_{*})$ as the Heaviside function.}
\label{fig1}
\end{figure}
Then modes in \eqref{eq:139} can be written as 
\begin{eqnarray}
    \label{eq:177i}
    \phi_{\omega>m}\approx \frac{1}{\sqrt{2\omega}}e^{-i\omega v+i\varphi_{\omega}/2}+\int \limits_{m}^{+\infty}\dfrac{d\omega^{\prime}}{2\pi}\dfrac{1}{\sqrt{2\omega^{\prime}}}\gamma(\omega,\omega^{\prime})e^{-i\omega^{\prime}u}+\int \limits_{m}^{+\infty}\dfrac{d\omega^{\prime}}{2\pi}\dfrac{1}{\sqrt{2\omega^{\prime}}}\gamma(\omega,-\omega^{\prime})e^{i\omega^{\prime}u},
\end{eqnarray}
which is valid for $r>R(t)$ and $\left|r-R(t)\right| \ll r_{g}$, as $t\rightarrow +\infty$.

Knowing the modes near the shell \eqref{eq:177i}, we can find the WKB modes outside the shell for arbitrary $r>R(t)$. In fact, assuming that for $r>R(t)$ the modes have the WKB form: 
\begin{gather}
   \phi_{\omega>m}(t,r_{*})=\label{eq:181}\\ =\dfrac{C}{\sqrt{2k(r_{*})}}e^{-i\omega t-i\int \limits_{R_{0*}}^{r_{*}}dxk(x)}
    +\int \limits_{m}^{+\infty}\dfrac{d\omega^{\prime}}{2\pi}\dfrac{\alpha(\omega,\omega^{\prime})}{\sqrt{2k^{\prime}(r_{*})}}e^{-i\omega^{\prime}t+i\int \limits_{R_{0*}}^{r_{*}}dxk^{\prime}(x)}+\int\limits_{m}^{+\infty}\dfrac{d\omega^{\prime}}{2\pi}\dfrac{\beta(\omega,\omega^{\prime})}{\sqrt{2k^{\prime}(r_{*})}}e^{i\omega^{\prime}t-i\int \limits_{R_{0*}}^{r_{*}}dx k^{\prime}(x)},
    \notag
\end{gather}
where $k^{\prime}(r_{*})=\sqrt{ \omega^{\prime \ 2}-m^{2}\left(1-\dfrac{r_{g}}{r(r_{*})}\right)}$, we find that near the horizon they behaves as
\begin{eqnarray}
    \label{eq:182}
   \phi_{\omega>m}(r\sim r_{g})=\dfrac{C}{\sqrt{2\omega}}e^{-i\omega v+i\omega R_{0*}}
    +\int \limits_{m}^{+\infty}\dfrac{d\omega^{\prime}}{2\pi}\dfrac{\alpha(\omega,\omega^{\prime})}{\sqrt{2\omega^{\prime}}}e^{-i\omega^{\prime}u-i\omega^{\prime}R_{0*}}+\int \limits_{m}^{+\infty}\dfrac{d\omega^{\prime}}{2\pi}\dfrac{\beta(\omega,\omega^{\prime})}{\sqrt{2\omega^{\prime}}}e^{i\omega^{\prime}u+i\omega^{\prime}R_{0*}}.
\end{eqnarray}
Comparing \eqref{eq:177i} and \eqref{eq:182} we get that 
\begin{gather}
    \label{eq:183}
  \left\{  \begin{matrix}
    C=e^{-i\omega R_{0*}}e^{i\varphi_{\omega }/2},\\
    \alpha(\omega ,\omega ^{\prime})=e^{i\omega ^{\prime}R_{0*}}\gamma(\omega ,\omega ^{\prime}), \\
    \beta(\omega ,\omega ^{\prime})=e^{-i\omega ^{\prime}R_{0*}}\gamma(\omega ,-\omega ^{\prime}),
    \end{matrix}\right.
\end{gather}
with $\gamma(\omega ,\omega ^{\prime})$ defined in \eqref{eq:180}.

As $t \rightarrow +\infty$ these harmonics behave as: 
 \begin{gather}
    \label{eq:185}
    \phi_{\omega>m}\approx
    \left\{
    \begin{matrix}
    \dfrac{e^{i\varphi_{\omega }/2}}{\sqrt{2\omega}}e^{-i\omega v}+{\displaystyle \int \limits_{m}^{+\infty}}\dfrac{d\omega ^{\prime}}{2\pi}\dfrac{\gamma(\omega,\omega^{\prime})}{\sqrt{2\omega^{\prime}}}e^{-i\omega^{\prime}u}+{\displaystyle \int \limits_{m}^{+\infty}}\dfrac{d\omega ^{\prime}}{2\pi}\dfrac{\gamma(\omega,-\omega^{\prime})}{\sqrt{2\omega^{\prime}}}e^{i\omega^{\prime}u}, \ \  r\rightarrow r_{g},\\
    \\
    \dfrac{e^{i\varphi_{\omega }/2}}{\sqrt{2k}}e^{-i\omega t-ikr_{*}}+{\displaystyle \int \limits_{m}^{+\infty}}\dfrac{d\omega ^{\prime}}{2\pi}\dfrac{\gamma(\omega ,\omega ^{\prime})}{\sqrt{2k^{\prime}}}e^{-i\omega ^{\prime}t+ik^{\prime}r_{*}}+{\displaystyle \int \limits_{m}^{+\infty}}\dfrac{d\omega ^{\prime}}{2\pi}\dfrac{\gamma(\omega ,-\omega ^{\prime})}{\sqrt{2k^{\prime}}}e^{i\omega ^{\prime}t-ik^{\prime}r_{*}}, \ \  r \rightarrow \infty,
    \end{matrix}\right.
\end{gather}
where 
 $k^{2}=\omega ^{2}-m^{2}$, and at $r \rightarrow +\infty$ we use the behaviour of $k(r_{*})$, depicted in  the fig.\ref{fig1}.
 
 It is worth to mention that we reexpand the harmonics in \eqref{eq:139} in terms of exponentials for convenience only, which simplify the calculations of the expectation value of the stress energy tensor, and clearly gives the Hawking radiation as we show below.

 Harmonics \eqref{eq:185} are defined for $\omega>m$. However, near the shell there are harmonics with $0<\omega<m$ which exponentially decay at spatial infinity, fig.\ref{fig2dm} 
 . They play an important role only in the vicinity of the shell, and have no contribution at spatial infinity. 
 
To find those harmonics we need to perform the following steps. First, near the shell the mass term vanishes, i.e. we need to glue linear combination of $e^{i\omega v}$ and $e^{i\omega u}$ with the solution of equations of motion inside the shell at $r=R(t)$. Second, we need to find such a linear combination that decays exponentially as $r \rightarrow +\infty$, i.e. $e^{-\sqrt{m^{2}-\omega^{2}}r_{*}}$. These two conditions complicate  the problem of finding the harmonics. We discuss the exact form of the harmonics in the appendix \ref{modesw<m}. Here we formulate the results only\footnote{Truly speaking we should have the sum over discrete spectrum for $\omega<m$. However, to simplify calculation of the integrals, we formally replace the sum with the integral over the interval $\omega<m$.}:
\begin{eqnarray}
\label{eq2}
\phi_{\omega<m}\approx 
    \int \limits_{\mu<|\omega^{\prime}|<m}\dfrac{d\omega^{\prime}}{2\pi}\gamma(w,\omega^{\prime})\sqrt{\frac{2}{\omega^{\prime}}}e^{-i\omega^{\prime }t}\sin{\left(-\omega^{\prime}r_{*}\right)} , \qquad   r>R(t), \qquad r \rightarrow r_{g},
\end{eqnarray}
with 
\begin{eqnarray}
\label{eq8}
\gamma(\omega,\omega^{\prime})= -2ir_{g}\dfrac{\sqrt{2|\omega^{\prime}|}}{\sqrt{2\omega}}e^{\pi \omega^{\prime}r_{g}}e^{2i\omega^{\prime}r_{g}\log{\omega r_{g}}}e^{-i\omega^{\prime}r_{g}}e^{-i\omega\tilde{A}}\Gamma(-2i\omega^{\prime}r_{g})\sin{\left(\omega_{-}r_{g}\right)}e^{i\pi/4},
\end{eqnarray}
where $\tilde{A}=r_{g}\sqrt{1-\frac{r_{g}}{R_{0}}}$.
The cut off $\mu \rightarrow 0$ comes from the demand of the validity of the WKB approximation, i.e. when $r \rightarrow r_{g}$ this condition is satisfied as long as the denominator of the equation 
\begin{equation}
    \label{eq4}
    \dfrac{1}{2mr_{g}}\dfrac{1}{\left[\omega^{2}/m^{2}-\left(1-r_{g}/r\right)\right]^{3/2}}\dfrac{r^{2}_{g}}{r^{2}}\left[1-\dfrac{r_{g}}{r}\right] \ll 1,
\end{equation}
is not equal to zero.

\section{The calculation of the covariantly conserved stress energy tensor}
\subsection{The stress energy tensor as \texorpdfstring{$t,r\rightarrow +\infty$}{TEXT}}\label{SETafter}

In this section we calculate the expectation value of the stress energy tensor using covariant point-splitting regularization at infinity $r\rightarrow +\infty$ as $t \rightarrow +\infty$. We perform calculations for two dimensional metric
\begin{equation}
    \label{eq:19}
    ds^{2}=C\left(u,v\right)dudv, \qquad \text{with} \qquad C\left(u,v\right)=C\left(u-v\right)=1-\dfrac{r_{g}}{r(r_{*})},
\end{equation}
with $u$,$v$ as in \eqref{que:13}.
For such a metric we have only two nonzero Christofel symbols
\begin{equation}
    \label{eq:21}
    \Gamma_{uu}^{u}=\partial_{u}\log{C}=\Gamma_{u}; \qquad \Gamma_{vv}^{v}=\partial_{v}\log{C}=\Gamma_{v}.
\end{equation}
The expectation value of
the stress energy tensor is calculated by the expansion of the expression
\begin{eqnarray}
    \label{eq:120}
     T_{\mu\nu}=\dfrac{1}{2}\langle \partial_{\alpha}\phi(x^{+})\partial_{\beta}\phi(x^{-})+\partial_{\beta}\phi(x^{-})\partial_{\alpha}\phi(x^{+})\rangle\left(e^{+\alpha}_{\mu}e^{-\beta}_{\nu}-\dfrac{1}{2}g_{\mu\nu}g^{\sigma\rho} e^{+\alpha}_{\sigma}e^{-\beta}_{\rho}\right)+\nonumber\\
    +\dfrac{m^{2}}{2}g_{\mu\nu}\dfrac{1}{2}\langle\phi(x^{+})\phi(x^{-})+\phi(x^{-})\phi(x^{+})\rangle,
\end{eqnarray}
in powers of $\epsilon$. Here the points $x^{\pm}$ lie on a geodesic passing through the point $x$ of interest, each at a proper distance $\epsilon$, but in opposite directions from $x$; $e^{\pm\alpha}_{\mu}$ -- are the matrices of the parallel transport along the geodesic from $x$ to $x^{\pm}$. In the appendix \ref{Calculations} we show that the expectation value of the stress energy tensor for the modes at $r \rightarrow \infty$ and $t \rightarrow +\infty$ (equation \eqref{eq:185}) has the form
\begin{equation}
    \label{eqU:45}
    T_{\mu\nu}=-\left[\dfrac{1}{4\pi \epsilon^{2} (t_{\alpha}t^{\alpha})}+\dfrac{R}{24 \pi}+\dfrac{m^{2}}{4\pi}\right]\left[\dfrac{t_{\mu}t_{\nu}}{t_{\alpha}t^{\alpha}}-\dfrac{1}{2}g_{\mu\nu}\right]+\Theta_{\mu\nu}-\dfrac{m^{2}}{4\pi}g_{\mu\nu}\left[\log{\left(\dfrac{m\sigma}{2}\right)}+\gamma\right],
\end{equation}
where 
\begin{gather}
    \label{eqU:46}
  \left\{  \begin{matrix}
    \Theta_{uu}=-\dfrac{1}{12\pi} C^{1/2}\partial^{2}_{u}C^{-1/2}+\dfrac{1}{8\pi}{\displaystyle \int \limits_{m}^{+\infty} }d\omega \dfrac{\omega}{e^{4\pi \omega r_{g}}-1}\left[\dfrac{\omega}{k}\left(1+\dfrac{k}{\omega}\right)^{2}\right],\\
    \Theta_{vv}=-\dfrac{1}{12\pi} C^{1/2}\partial^{2}_{v}C^{-1/2}+\dfrac{1}{8\pi}{\displaystyle \int \limits _{m}^{+\infty}}d\omega \dfrac{\omega}{e^{4\pi \omega r_{g}}-1}\left[\dfrac{\omega}{k}\left(1-\dfrac{k}{\omega}\right)^{2}\right],\\
    \Theta_{vu}=\Theta_{uv}=0,
    \end{matrix}\right.
\end{gather}
with  $\sigma^{2}\equiv C(u,v)\left(u^{+}-u^{-}\right)\left(v^{+}-v^{-}\right)\rightarrow{}0$ being the geodesic distance between the points $x^{+}$ and $x^{-}$, and $\gamma$ is the Euler–Mascheroni constant.

Following \cite{Davies:1976ei}, we neglect terms of the form $t_{\mu}t_{\nu}$, and get 
\begin{equation}
    \label{que:17}
    T_{\mu\nu}=\Theta_{\mu\nu}+\dfrac{1}{48 \pi}Rg_{\mu \nu}-\dfrac{m^{2}}{4\pi}g_{\mu \nu}\left[\log{\left(\dfrac{m\sigma}{2}\right)}+\gamma-\dfrac{1}{2}\right],
\end{equation}
which is covariantly conserved:
\begin{equation}
    \label{eq:47}
    \nabla_{\mu}T^{\mu \nu}=0.
\end{equation}
Finally, the flux at infinity at the final stage of the collapse has the form
\begin{eqnarray}
    \label{eq:199}
    T_{tr_{*}}=T_{uu}-T_{vv}=\nonumber\\
    =\dfrac{1}{8\pi}\int \limits_{m}^{+\infty}d\omega \dfrac{\omega}{e^{4\pi \omega r_{g}}-1}\left[\dfrac{\omega}{k}\left(1+\dfrac{k}{\omega}\right)^{2}\right]-\dfrac{1}{8\pi}\int \limits_{m}^{+\infty}d\omega \dfrac{\omega}{e^{4\pi \omega  r_{g}}-1}\left[\dfrac{\omega}{k}\left(1-\dfrac{k}{\omega}\right)^{2}\right]=\nonumber\\
    =\int \limits_{m}^{+\infty}\dfrac{d\omega}{2\pi} \dfrac{\omega}{e^{4\pi \omega r_{g}}-1}, \qquad r,t\rightarrow +\infty.
\end{eqnarray}
Even though the results were obtained in the consideration of heavy scalar field, $mr_{g} \gg 1$, still, if we, formally, set $m=0$, we reproduce the result of the paper \cite{Davies:1976ei}.

The expectation value of the stress energy tensor \eqref{que:17} has the divergent logarithmic term $\log{\left(\epsilon m\right)}$ which necessarily needs to be renormalized. We assume that the renormalization of \eqref{que:17} is the same as in \cite{Bunch:1978aq} (see also \cite{WaldCommun.Math}-
\cite{Mazzitelli:2011st}). Moreover, in this paper the main goal was to find the flux near the horizon and at spatial infinity which are not affected by the logarithmic term. 

\subsection{The stress energy tensor as \texorpdfstring{$t\rightarrow +\infty$}{TEXT} in the vicinity of the shell}\label{SETbefore}

Calculation of the expectation value of the stress energy tensor in the vicinity of the shell with the harmonics \eqref{eq:185} 
is very similar to the one presented in the previous section and appendix \ref{Calculations}. However, there is also contribution from harmonics \eqref{eq2}, i.e. from $\omega<m$ part of the spectrum. In the appendix \ref{SEThorizon} we provide details of the calculation, and here we formulate the result only:
\begin{equation}
    \label{que:20}
    T_{\mu\nu}=\Theta^{\prime}_{\mu\nu}+\dfrac{1}{48 \pi}Rg_{\mu \nu}-\dfrac{m^{2}}{4\pi}g_{\mu \nu}\left[\log{\left(m\sigma\right)}+\gamma-\dfrac{1}{2}\right],
\end{equation}
with
\begin{gather}
    \label{que:21}
  \left\{  \begin{matrix}
    \Theta^{\prime}_{uu}=-\dfrac{1}{12\pi} C^{1/2}\partial^{2}_{u}C^{-1/2}+{\displaystyle \int \limits _{0}^{+\infty} }\dfrac{d\omega}{2\pi} \dfrac{\omega}{e^{4\pi \omega r_{g}}-1},\\
    \Theta^{\prime}_{vv}=-\dfrac{1}{12\pi} C^{1/2}\partial^{2}_{v}C^{-1/2}+{\displaystyle \int \limits _{0}^{m} }\dfrac{d\omega}{2\pi} \dfrac{\omega}{e^{4\pi \omega r_{g}}-1},\\
    \Theta^{\prime}_{vu}=\Theta^{\prime}_{uv}=0,
    \end{matrix}\right.
\end{gather}
which is covariantly conserved:
\begin{equation}
    \label{eq:47ryrhfr}
    \nabla_{\mu}T^{\mu \nu}=0.
\end{equation}
Using that 
\begin{eqnarray}
    \label{que:22}
    -\dfrac{1}{12\pi} C^{1/2}\partial^{2}_{u}C^{-1/2}=-\dfrac{1}{12\pi} C^{1/2}\partial^{2}_{v}C^{-1/2}=-\dfrac{1}{192 \pi r_{g}^{2}}   \qquad \qquad r\rightarrow{r_{g}},
\end{eqnarray}
and the fact that
\begin{eqnarray}
    \label{que:23}
    \int \limits_{0}^{+\infty}\dfrac{d\omega}{2\pi} \dfrac{\omega}{e^{4\pi \omega r_{g}}-1}=\dfrac{1}{192\pi r_{g}^{2}},
\end{eqnarray}
we obtain 
\begin{gather}
    \label{que:24}
   \left\{ \begin{matrix}
     T_{vv}\approx -\int \limits_{m}^{+\infty}\dfrac{d\omega}{2\pi} \dfrac{\omega}{e^{4\pi \omega r_{g}}-1} , \\
     T_{uu} \approx 0,
    \end{matrix}\right. \qquad \qquad r\rightarrow r_{g}.
\end{gather}
When $m=0$ we reproduce exactly \eqref{eq:58}. Also, the fact that  $T_{uu}=0$ as $r \rightarrow r_{g}$ is in full agreement with the the results of \cite{Christensen:1977jc}.

\section{Conclusions}\label{Conclusions}
First, we have found the behaviour of the semiclassical harmoinics for the massive scalar field outside the shell as $t \rightarrow +\infty$ that are given in \eqref{eq:177i} and \eqref{eq2}. These are the harmonics which diagonalize the free Hamiltonian before the collapse.

Second, we find the expectation value of the stress energy tensor \eqref{eq:120} with the help of the covariant point splitting regularization in the vicinity of the shell
\begin{gather}
    \label{cs:8}
  \left\{  \begin{matrix}
     T_{uu}\approx 0, \\
     T_{vv} \approx-{\displaystyle \int \limits_{m}^{+\infty} }\dfrac{d\omega}{2\pi} \dfrac{\omega}{e^{4\pi \omega r_{g}}-1},
    \end{matrix}\right. \qquad \qquad r\rightarrow r_{g} \qquad \text{and} \qquad t \rightarrow +\infty.
\end{gather}
Similarly to the massless case there is a negative flux near the shell. 

Third, we have found that far away from the shell the flux is
\begin{gather} 
    \label{cs:12}
  \left\{  \begin{matrix}
    T_{uu} \approx 
    \dfrac{1}{8\pi}{\displaystyle \int \limits_{m}^{+\infty}}d\omega \dfrac{\omega}{e^{4\pi \omega r_{g}}-1}\left[\dfrac{\omega}{k}\left(1+\dfrac{k}{\omega}\right)^{2}\right], \\
    T_{vv} \approx \dfrac{1}{8\pi}{\displaystyle \int \limits_{m}^{+\infty}}d\omega \dfrac{\omega}{e^{4\pi \omega r_{g}}-1}\bigg[\dfrac{\omega}{k}\left(1-\dfrac{k}{\omega}\right)^{2}\bigg],
    \end{matrix}\right.  \qquad r,t \rightarrow +\infty.
\end{gather}
Furthermore, the total flux $T_{tr_{*}}=T_{uu}-T_{vv}$ is the same in the vicinity of the shell and at the spatial infinity and is equal to
\begin{eqnarray}
    \label{cs:14} 
    T_{tr_{*}} \approx \int \limits_{m}^{+\infty}\dfrac{d\omega }{2\pi}\dfrac{\omega }{e^{4\pi \omega r_{g}}-1}.
\end{eqnarray}

It is also interesting to point out that in the paper \cite{Akhmedov:2015xwa} the flux was calculated in the vicinity of the collapsing shell in four dimensions. The result had the form
\begin{gather} 
    \label{conc:1}
   \left\{ \begin{matrix}
         J_{v}\sim {\displaystyle \int \limits_{S^{2}} } \sin{\theta}d\theta d\varphi \langle T_{vv}\rangle \sim \dfrac{1}{2}{\displaystyle \int \limits_{m}^{+\infty} }d\omega  \omega , \\
         J_{u}\sim  {\displaystyle \int \limits_{S^{2}} } \sin{\theta}d\theta d\varphi \langle T_{uu}\rangle \sim \sqrt{1-r_{g}/R_{0}}{\displaystyle \int \limits_{m}^{+\infty} }\dfrac{d\omega }{2\pi}\dfrac{\omega }{e^{4\pi \omega r_{g}}-1} +\dfrac{1}{2}{\displaystyle \int \limits_{m}^{+\infty}}d\omega  \omega ,
    \end{matrix}\right.
\end{gather}
where the usual point splitting regularization was used, since authors were interested in finding the ${\displaystyle \int _{S^{2} }} \sin{\theta}d\theta d\varphi \langle T_{tr_{*}}\rangle$ only. Even though they received correct thermal radiation still, with this method of regularization they did not observe negative $vv$- flux near the shell. The latter is seen if one uses the covariant point splitting method. It is present in the part of the stress energy tensor which has the form
 $$-\dfrac{1}{12\pi} C^{1/2}\partial^{2}_{u}C^{-1/2}.$$
 
Because we find the explicit form of the harmonics outside the shell for heavy fields, it will be interesting to calculate  the loop corrections to the stress–energy flux in  selfinteracting $\phi^{4}$
model. It is possible that in this model the  perturbative IR corrections grow
with time and, if one considers a long enough period of time they can even dominate over the tree level contribution, which is true in the model of massive scalar in the background of collapsing thin shell in four dimensions, \cite{Akhmedov:2015xwa}. There is a hope that in 2D it is possible to do the resummation of all the leading loop
corrections, which was not obtained in four dimensional model.

\section{Acknowledgements}
I would like to thank E.T. Akhmedov for proposing this problem to me and proof reading this
text. Also, I would like to acknowledge valuable discussions with E.T. Akhmedov, S. Alekseev, P. Anempodistov, L. Astrakhantsev, F. Popov  and D.Trunin. 

\newpage

\appendix
\section{Calculating the expectation value of the stress energy tensor at spatial infinity}\label{Calculations}
In this section we provide some details of calculations of $T_{\mu \nu}$ with the help of the covariant point splitting regularization (for the detailed discussion of this method see, for example, \cite{Davies covariant point splitting}-\cite{Akhmedov:2020ryq}).

We assume that the separation of points on geodesic is infinitesimal, which means that we can use the expansion of the form
\begin{equation}
    \label{eq:23}
    x^{\mu}(\pm\epsilon)=x^{\mu} \pm \epsilon t^{\mu}+\dfrac{1}{2}\epsilon^{2}a^{\mu}\pm \dfrac{1}{6}\epsilon^{3}b^{\mu}+... .
\end{equation}
We find coefficients $a^{\mu}, b^{\mu}, ...$ in \eqref{eq:23} by putting $x^{\mu}$ into geodesic equation:
\begin{equation}
    \label{eq:24}
    \dfrac{d^2x^{\mu}}{d\tau^{2}}+\Gamma^{\mu}_{\nu\lambda}\dfrac{dx^{\nu}}{d\tau}\dfrac{dx^{\lambda}}{d\tau}=0,
\end{equation}
with the initial conditions:
\begin{equation}
    \label{eq:25}
    x^{\mu}(\tau=0)=x^{\mu}, \qquad \dfrac{dx^{\mu}}{d\tau}(\tau=0)=t^{\mu}=(t^{u},t^{v}).
\end{equation}

The solution has the following form:
\begin{gather}
    \label{eq:27}
  \left\{  \begin{matrix}
     a^{\mu}=-\Gamma^{\mu}_{\nu\lambda}t^{\nu}t^{\lambda}, \\
      b^{\mu}=-\Gamma^{\mu}_{\nu \lambda}\left(a^{\nu}t^{\lambda}+t^{\nu}a^{\lambda}\right)-t^{s}\partial_{s}\Gamma^{\mu}_{\nu \lambda}t^{\nu}t^{\lambda},
    \end{matrix}\right.
\end{gather}
or explicitly in light-cone coordinates:
\begin{gather}
    \label{eq:29}
   \left\{ \begin{matrix}
    a^{u}=-\Gamma_{u}\left(t^{u}\right)^{2}, \\ a^{v}=-\Gamma_{v}\left(t^{v}\right)^{2},
    \end{matrix}\right. \qquad \text{and} \qquad
 \left\{   \begin{matrix}
    b^{u}=-\left(t^{u}\right)^{2} t^{v}\partial_{v}\Gamma_{u}+\left(2\Gamma^{2}_{u}-\partial_{u}\Gamma_{u}\right)\left(t^{u}\right)^{3}, \\
     b^{v}=-\left(t^{v}\right)^{2} t^{u}\partial_{u}\Gamma_{v}+\left(2\Gamma^{2}_{v}-\partial_{v}\Gamma_{v}\right)\left(t^{v}\right)^{3}.
    \end{matrix}\right.
\end{gather}
Next, the matrix of the parallel transport has the form:
\begin{equation}
    \label{eq:30}
    \dfrac{de^{\mu}_{\nu}}{d\tau}+\Gamma^{\mu}_{\rho\sigma}\dfrac{dx^{\rho}}{d\tau}e^{\sigma}_{\nu}=0,
\end{equation}
with the initial condition: $e^{\mu}_{\nu}(\tau=0)=\delta^{\mu}_{\nu}$.
The solution has the form:
\begin{equation}
    \label{eq:31}
    e(\tau)^{\mu}_{\nu}=\delta^{\mu}_{\nu}+\tau t^{\mu}_{\nu}+\dfrac{1}{2}\tau^{2} a^{\mu}_{\nu}+... .
\end{equation}
with
\begin{gather}
    \label{eq:32}
  \left\{  \begin{matrix}
     t^{\mu}_{\nu}=-\Gamma^{\mu}_{\rho\nu}t^{\rho}, \\
     a^{\mu}_{\nu}=\Gamma^{\mu}_{\rho\nu}\Gamma^{\rho}_{\alpha\beta}t^{\alpha}t^{\beta}+\Gamma^{\mu}_{\rho\sigma}\Gamma^{\sigma}_{\alpha\nu}t^{\rho}t^{\alpha}-t^{\alpha}t^{\rho}\partial_{\alpha}\Gamma^{\mu}_{\rho\nu}.
    \end{matrix}\right.
\end{gather}
As can be seen from the equations above, the matrix $e^{\mu}_{\nu}$ is diagonal.

For completeness, we again write the explicit form of the harmonics as $t \rightarrow \infty$ and $r \rightarrow +\infty$: 
\begin{eqnarray}
    \label{eq:186}
    \phi_{\omega}(t,r_{*})=\dfrac{e^{i\varphi_{\omega}/2}}{\sqrt{2k}}e^{-i\omega t-ikr_{*}}+\int \limits_{m}^{+\infty}\dfrac{d\omega^{\prime}}{2\pi}\dfrac{\gamma(\omega,\omega^{\prime})}{\sqrt{2k^{\prime}}}e^{-i\omega^{\prime}t+ik^{\prime}r_{*}}+\int\limits_{m}^{+\infty}\dfrac{d\omega^{\prime}}{2\pi}\dfrac{\gamma(\omega,-\omega^{\prime})}{\sqrt{2k^{\prime}}}e^{i\omega^{\prime}t-ik^{\prime}r_{*}},
\end{eqnarray}
where $\gamma(\omega,\omega^{\prime})$ is defined in \eqref{eq:180}. 

The integral over $\omega$ in the expression for 
$\langle \partial_{u}\hat{\phi}(x^{+})\partial_{u}\hat{\phi}(x^{-})\rangle$ can first be evaluated as follows
\begin{gather}
    \label{eq:189} 
    \int\limits_{m}^{+\infty}\dfrac{d\omega}{2\pi}\gamma(\omega,\omega^{\prime})\gamma^{*}(\omega,\omega^{\prime\prime})
    =\nonumber\\
    =16r^{2}_{g}\sqrt{|\omega^{\prime}||\omega^{\prime\prime}|}e^{\pi r_{g}(\omega^{\prime}+\omega^{\prime\prime})} \Gamma(-2i\omega^{\prime}r_{g}) \Gamma(2i\omega^{\prime\prime}r_{g})e^{-ir_{g}(\omega^{\prime}-\omega^{\prime\prime})}\int\limits_{m}^{+\infty}\dfrac{d\omega}{2\pi}\dfrac{e^{2ir_{g}(\omega^{\prime}-\omega^{\prime\prime})\log(\omega r_{g})}}{\omega}\sin^{2}{(\omega_{-}r_{g})} = \\ \nonumber
   = 16r^{2}_{g}\sqrt{|\omega^{\prime}||\omega^{\prime\prime}|}e^{\pi r_{g}(\omega^{\prime}+\omega^{\prime\prime})} \Gamma(-2i\omega^{\prime}r_{g}) \Gamma(2i\omega^{\prime\prime}r_{g})e^{-ir_{g}(\omega^{\prime}-\omega^{\prime\prime})}\left[\dfrac{1}{8r_{g}}\delta(\omega ^{\prime}-\omega ^{\prime\prime}) \right] + \text{regular terms},
\end{gather}
where one can show that the contribution from "regular terms" are negligible as $t \rightarrow +\infty$.
In all, we find that:
\begin{eqnarray}
    \label{eq:193}
    \langle \partial_{u}\hat{\phi}(x^{+})\partial_{u}\hat{\phi}(x^{-})\rangle=\nonumber\\
    =\dfrac{1}{16\pi}\int\limits_{m}^{+\infty}d\omega \dfrac{(\omega -k)^{2}}{8k}e^{-i\omega (t^{+}-t^{-})-ik(r_{*}^{+}-r_{*}^{-})}
    +\dfrac{1}{16\pi}\int\limits_{m}^{+\infty}d\omega  \dfrac{(\omega +k)^{2}}{k}e^{-i\omega (t^{+}-t^{-})+ik(r^{+}_{*}-r^{-}_{*})}+\nonumber\\
    +\dfrac{1}{8\pi}\int\limits_{m}^{+\infty}d\omega  \dfrac{\omega }{e^{4\pi \omega  r_{g}}-1}\left[\dfrac{\omega }{k}\left(1+\dfrac{k}{\omega }\right)^{2}\right]+\text{suppressed terms}.
\end{eqnarray}
The integrals in \eqref{eq:193} can be calculated using the table integrals from \cite{Gradshteyn} . Hence, the result is:
\begin{eqnarray}
    \label{eq:196}
    T_{uu}=\langle \partial_{u}\hat{\phi}(x^{+})\partial_{u}\hat{\phi}(x^{-})\rangle e^{+u}_{u}e^{-u}_{u}=\nonumber\\
    =-\left[\dfrac{1}{4\pi \epsilon^{2} (t_{\alpha}t^{\alpha})}+\dfrac{R}{24 \pi}+\dfrac{m^{2}}{4\pi}\right]\dfrac{t_{u}t_{u}}{t_{\alpha}t^{\alpha}}+\Theta_{uu}+\text{suppresed terms},
\end{eqnarray}
where
\begin{equation}
    \label{eq:197}
    \Theta_{uu}=-\dfrac{1}{12\pi} C^{1/2}\partial^{2}_{u}C^{-1/2}+\dfrac{1}{8\pi}\int\limits_{m}^{+\infty}d\omega  \dfrac{\omega }{e^{4\pi \omega  r_{g}}-1}\left[\dfrac{\omega }{k}\left(1+\dfrac{k}{\omega }\right)^{2}\right].
\end{equation}
Similarly, 
\begin{equation}
\label{eq:197v}
T_{vv}=-\left[\dfrac{1}{4\pi \epsilon^{2} (t_{\alpha}t^{\alpha})}+\dfrac{R}{24 \pi}+\dfrac{m^{2}}{4\pi}\right]\dfrac{t_{v}t_{v}}{t_{\alpha}t^{\alpha}}+\Theta_{vv},
\end{equation}
where
\begin{equation}
    \label{eq:198v}
    \Theta_{vv}=-\dfrac{1}{12\pi} C^{1/2}\partial^{2}_{v}C^{-1/2}+\dfrac{1}{8\pi}\int\limits_{m}^{+\infty}d\omega  \dfrac{\omega }{e^{4\pi \omega  r_{g}}-1}\left[\dfrac{\omega }{k}\left(1-\dfrac{k}{\omega }\right)^{2}\right].
\end{equation}
Finally, the calculation of $T_{uv}$ component is very similar. The result has the form:
\begin{eqnarray}
    \label{eq:135}
    T_{vu}\approx-\dfrac{m^{2}}{4\pi}g_{vu}\bigg[\log{\bigg(\dfrac{m\sigma}{2}\bigg)}+\gamma\bigg],
\end{eqnarray}
where $\gamma$ is the Euler–Mascheroni constant.

\section{The form of harmonics with \texorpdfstring{$0<\omega<m$}{TEXT} as  \texorpdfstring{$r \rightarrow r_{g}$}{TEXT} }\label{modesw<m}
\begin{figure}[t!]
\centering
\begin{tikzpicture}
\begin{axis}[
	axis x line=center,
	axis y line=center,
	domain=-15:15,
	xlabel=$r_{*}$, xlabel style={at=(current axis.right of origin), anchor= west},
	ylabel=$V(r_{*})$, ylabel style={at=(current axis.above origin), anchor=south },
	samples=100,
	ymin=-0.01,ymax=200,xmin=-15,xmax=15, xtick={0.01,3},xticklabels={$0$,$r_{*}^{tn}(\omega)$}, ytick=\empty]
	  \addplot+[mark=none, draw=blue, thick] {atan(x-2) + 89};
    

    \draw [black, dashed, mark size=3 pt] (axis cs:+50,+177) -- (axis cs: 0,+177) node [left]{$m^{2}$};
    \draw [black, mark size=3 pt] (axis cs:10,+135) -- (axis cs: -85,+135) node [above]{$\omega^{2}$};
    \draw [red, dashed, mark size=3 pt] (axis cs:3,+0) -- (axis cs: 3,+160) node [above right]{$$};
\end{axis}
\node [color=black, right]at (1.6,4.2) {$\omega^{2}$};
\end{tikzpicture}
\caption{The form of the potential $V[r_{*}]$ in \eqref{eq:146}. The point $r^{tn}_{*}(\omega)$ is the turning point which separates classically allowed and forbidden regions.}
\label{fig2dm}
\end{figure}
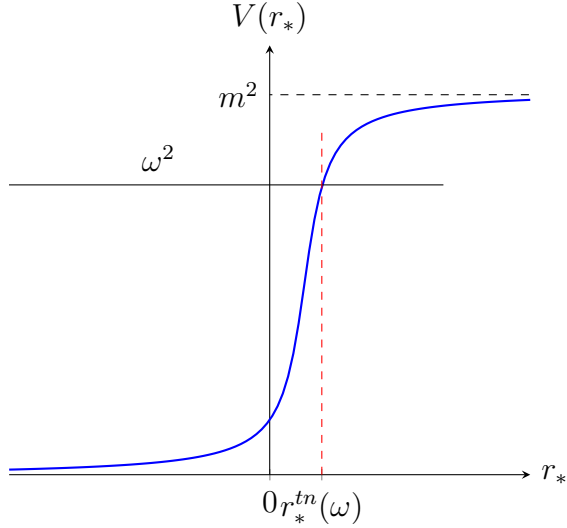

The quasiclassical harmonics with $\omega<m$ from \eqref{eq:146}  have the form \cite{Migdal}
\begin{eqnarray}
    \label{d24}
     \phi_{\omega} \approx \left\{
    \begin{matrix}
    \dfrac{D}{\sqrt{k(r_{*})}}e^{-i\omega t}\sin{\left(\int\limits_{r_{*}}^{r^{tn}_{*}(\omega)}dxk(x)+\dfrac{\pi}{4}\right)}, \qquad r_{*}< r_{*}^{tn}(\omega), \ \  \\
    \dfrac{D}{2\sqrt{|k(r_{*})|}}e^{-i\omega t}e^{-\int\limits_{r^{tn}_{*}(\omega)}^{r_{*}}dx|k(x)|}, \qquad \qquad r_{*}> r_{*}^{tn}(\omega), \ \
    \end{matrix}
    \right. 
\end{eqnarray}
with $V\left[r_{*}\right]=        m^{2}\left(1-\dfrac{r_{g}}{r[r_{*}]}\right)$ 
,  $k(r_{*})=\sqrt{\omega^{2}-V[r_{*}]}$ , and the turning point $r^{tn}_{*}(\omega)$ is as follows (see fig. \ref{fig2dm})
\begin{eqnarray}
    \label{d25}
   \omega^{2}=V[r^{tn}_{*}], \qquad \text{hence} \qquad r^{tn}_{*}(\omega)=r_{g}\dfrac{m^{2}}{m^{2}-\omega^{2}}+r_{g}\log{\dfrac{\omega^{2}}{m^{2}-\omega^{2}}}.
\end{eqnarray}

Now we discuss how to glue the harmonics at the shell. Particularly, we suggest that at the late stage of collapse, i.e. $t\rightarrow +\infty$ the behaviour of the harmonics inside the shell is the same as one before collapse, with the change $t_{-}=\sqrt{1-\frac{r_{g}}{R_{0}}}t  \ \Longrightarrow \  t_{-}\approx\frac{R_{0}-r_{g}}{c}\left(1-e^{-\frac{t}{r_{g}}}\right)$:
\begin{equation}
\label{eq1}   
\phi_{\omega}(t,r)= \dfrac{1}{\sqrt{2k(R_{0})}}e^{-iw_{-}t_{-}}\left(e^{ik(R_{0})\hat{r}}-e^{-ik(R_{0})\hat{r}}\right), \quad  r<R(t), \qquad m\sqrt{1-r_{g}/R_{0}}<\omega<+\infty,
\end{equation}
where $w_{-}=\dfrac{\omega}{\sqrt{1-r_{g}/R_{0}}}$,  \ \ $k(R_{0})=\sqrt{\omega^{2}-m^{2}\left(1-\frac{r_{g}}{R_{0}}\right)}$, $\hat{r}=\frac{r}{\sqrt{1-\frac{r_{g}}{R_{0}}}}$. 
However, in such an approximation there is a problem. As $t \rightarrow +\infty$ outside the shell we have the oscillating harmonics with $0<\omega<m$ when $r \rightarrow r_{g}$, while inside the shell (see  \eqref{eq1}) we have oscillating harmonics only for $m\sqrt{1-r_{g}/R_{0}}<\omega<m$. To glue the harmonics at the shell we need to make the following approximation: $k(R_{0})=\sqrt{\omega^{2}-m^{2}\left(1-\frac{r_{g}}{R_{0}}\right)} \rightarrow \omega$  which is fair since we assume that $R_{0} \approx r_{g}$. In such a way we can make the oscillating spectrum inside the shell in the region $0<\omega<m$. And after such a procedure we can glue the harmonics on the surface of the shell. In other words we can write the harmonics with $\omega<m$ as 
\begin{eqnarray}
\label{eq!!2}
\phi_{\omega}\approx \left\{ \begin{matrix}
    -2i\dfrac{e^{-i\omega _{-}t_{-}}}{\sqrt{2\omega}}\sin{\omega_{-}r}, & r<R(t), \\
    \int \limits_{\mu<|\omega^{\prime}|<m}\dfrac{d\omega^{\prime}}{2\pi}\gamma(w,\omega^{\prime})\sqrt{\frac{2}{\omega^{\prime}}}e^{-i\omega^{\prime }t}\sin{\left(-\omega^{\prime}r_{*}\right)} &  r>R(t), \ |R(t)-r_{g}| \ll r_{g},
    \end{matrix}\right.
\end{eqnarray}
Let us make a few comments about the form of the harmonics outside the shell. First of all, we approximate 
\begin{eqnarray}
\label{eq3}
\sqrt{\dfrac{2}{k(r_{*})}}\sin{\left(\int \limits_{r_{*}}^{r^{tn}_{*}(\omega)}dx\sqrt{\omega^{2}-V[x]}+\frac{\pi}{4}\right)} \approx \sqrt{\frac{2}{\omega}}\sin{\left(-\omega r_{*}\right)}, \ \ r \rightarrow r_{g},
\end{eqnarray}
where we assume\footnote{Without this assumption we can not drop the phase $\pi/4$ in \eqref{eq3}, which means that in \eqref{eq101!} we would have $\pi/2$ instead of $\pi$. But as we discuss in the appendix \ref{SEThorizon} the phase for $\mu<\omega<m$ is irrelevant for $T_{uv}$. It is also does not contribute neither to $T_{uu}$ nor to $T_{vv}$.} that $|\mu r_{*}| \gtrsim 1$ as $r_{*} \rightarrow -\infty$. Second, the constant $\mu \rightarrow 0$ comes from the validity of the WKB approach, i.e. when $r \rightarrow r_{g}$ this condition is satisfied as long as the denominator of the equation 
\begin{equation}
    \label{eq!!4}
    \dfrac{1}{2mr_{g}}\dfrac{1}{\left[\omega^{2}/m^{2}-\left(1-r_{g}/r\right)\right]^{3/2}}\dfrac{r^{2}_{g}}{r^{2}}\left[1-\dfrac{r_{g}}{r}\right] \ll 1,
\end{equation}
is not equal to zero. 
Now we are ready to glue the modes \eqref{eq!!2} on the surface of the shell, where $R(t) \approx r_{g}\left(1+\frac{R_{0}-r_{g}}{r_{g}}e^{-\frac{t}{r_{g}}} \right) \approx r_{g}$, and $R_{*}(t)\approx R_{0*}-t+\left(r_{g}-R_{0}\right)\left(1-e^{-\frac{t}{r_{g}}}\right) \approx R_{0*}-t$:
\begin{gather}
\label{eq5}
-\dfrac{2i}{\sqrt{2\omega}}e^{-i\omega\left(\tilde{B}e^{-t/r_{g}}+\tilde{A}\right)}\sin{\omega_{-}r_{g}} \approx \int \limits_{\mu<|\omega^{\prime}|<m}\dfrac{d\omega^{\prime}}{2\pi}\gamma(w,\omega^{\prime})\dfrac{1}{i}\sqrt{\frac{1}{2\omega^{\prime}}}e^{-i\omega^{\prime}t}
\left(e^{-i\omega^{\prime}\left(R_{0*}-t\right)}-e^{i\omega^{\prime}\left(R_{0*}-t\right)}\right),
\end{gather}
where $\tilde{B}=-r_{g}\sqrt{1-\frac{r_{g}}{R_{0}}}$, $\tilde{A}=-\tilde{B}$. Making the Fourier transformation in $t$, we find that 
\begin{eqnarray}
\label{eq6}
f(\omega, \omega^{\prime})=\left[\int \limits_{\mu<|\tilde{\omega}|<m}\dfrac{d\tilde{\omega}}{2\pi}\gamma(w, \tilde{\omega})\sqrt{\dfrac{1}{2\tilde{\omega}}}e^{-i\tilde{\omega}R_{0*}}\right]\delta(2\omega^{\prime})-\dfrac{1}{2}\gamma(\omega,\omega^{\prime})\sqrt{\dfrac{1}{2|\omega^{\prime}|}}e^{i\omega^{\prime}R_{0*}},
\end{eqnarray}
where 
\begin{eqnarray}
\label{eq7}
f(\omega,\omega^{\prime})=-\dfrac{2ir_{g}}{\sqrt{2\omega}}e^{\pi \omega^{\prime}r_{g}}e^{2i\omega^{\prime}r_{g}\log{\omega r_{g}}}e^{i\omega^{\prime}(R_{0*}-r_{g})}e^{-i\omega\tilde{A}}\Gamma(-2i\omega^{\prime}r_{g})\sin{\left(\omega_{-}r_{g}\right)}
\end{eqnarray}
Since in our approximation $\mu< |\omega^{\prime}|<m$ we can drop the term with delta function and in such a way find the Bogoliubov coefficients:
\begin{eqnarray}
\label{eq!8}
\gamma(\omega,\omega^{\prime})=\sqrt{2|\omega^{\prime}|}f(\omega,\omega^{\prime})e^{-i\omega^{\prime}R_{0*}}= \nonumber \\
=-2ir_{g}\dfrac{\sqrt{2|\omega^{\prime}|}}{\sqrt{2\omega}}e^{\pi \omega^{\prime}r_{g}}e^{2i\omega^{\prime}r_{g}\log{\omega r_{g}}}e^{-i\omega^{\prime}r_{g}}e^{-i\omega\tilde{A}}\Gamma(-2i\omega^{\prime}r_{g})\sin{\left(\omega_{-}r_{g}\right)}.
\end{eqnarray}

\section{Calculating the expectation value of the stress energy tensor near the shell}\label{SEThorizon}

As was mentioned before, the calculation of the expectation value of the stress energy tensor in the vicinity of the shell with the harmonics \eqref{eq:185} is very similar to the one presented in the previous section and appendix \ref{Calculations}. Hence, the result is:
\begin{eqnarray}
    \label{eq:196f}
    T^{\omega>m}_{uu}=\int \limits_{m}^{+\infty} \partial_{u}\phi_{\omega}(x^{+})\partial_{u}\phi_{\omega}(x^{-})\rangle e^{+u}_{u}e^{-u}_{u}=\nonumber\\
    =-\left[\dfrac{1}{4\pi \epsilon^{2} (t_{\alpha}t^{\alpha})}+\dfrac{R}{24 \pi}+\dfrac{m^{2}}{4\pi}\right]\dfrac{t_{u}t_{u}}{t_{\alpha}t^{\alpha}}+\Theta_{uu}+\text{suppresed terms},
\end{eqnarray}
where
\begin{equation}
    \label{eq:197f}
    \Theta_{uu}=-\dfrac{1}{12\pi} C^{1/2}\partial^{2}_{u}C^{-1/2}+\int\limits_{m}^{+\infty}\dfrac{d\omega }{2\pi} \dfrac{\omega }{e^{4\pi \omega  r_{g}}-1}-\dfrac{m^{2}}{8\pi}.
\end{equation}
Similarly, 
\begin{equation}
\label{eq:197vf}
T^{\omega>m}_{vv}=-\left[\dfrac{1}{4\pi \epsilon^{2} (t_{\alpha}t^{\alpha})}+\dfrac{R}{24 \pi}+\dfrac{m^{2}}{4\pi}\right]\dfrac{t_{v}t_{v}}{t_{\alpha}t^{\alpha}}+\Theta_{vv},
\end{equation}
where
\begin{equation}
    \label{eq:198fv}
    \Theta_{vv}=-\dfrac{1}{12\pi} C^{1/2}\partial^{2}_{v}C^{-1/2}-\dfrac{m^{2}}{8\pi}.
\end{equation}


Now we need to calculate the contribution to the expectation value of the stress energy tensor of the harmonics with $\mu<\omega<m$:
\begin{eqnarray}
\label{d27f}
\phi_{\omega<m}\approx 
    \int \limits_{\mu<|\omega^{\prime}|<m}\dfrac{d\omega^{\prime}}{2\pi}\gamma(w,\omega^{\prime})\sqrt{\frac{2}{\omega^{\prime}}}e^{-i\omega^{\prime }t}\sin{\left(-\omega^{\prime}r_{*}\right)} , \qquad   r>R(t), \qquad r \rightarrow r_{g},
\end{eqnarray}
with $\gamma(\omega,\omega^{\prime})$ defined in \eqref{eq!8}:
\begin{gather}
\label{eq9!}
T_{uu}=\int \limits_{\mu<|\omega^{\prime}|<m}\dfrac{d\omega^{\prime}}{2\pi}\int \limits_{\mu<|\omega^{\prime \prime}|<m}\dfrac{d\omega^{\prime \prime}}{2\pi}\dfrac{\omega^{\prime}\omega^{\prime \prime}}{\sqrt{2|\omega^{\prime}|2|\omega^{\prime \prime}|}} e^{-i\omega^{\prime}u_{+}+i\omega^{\prime\prime}u_{-}}\int \limits_{\mu}^{m}\dfrac{d\omega}{2\pi}\gamma(\omega,\omega^{\prime})\gamma^{*}(\omega,\omega^{\prime\prime}).
\end{gather}
Taking the integral over $\omega$ and isolating the leading contribution only we get:
\begin{gather}
\label{eq13!}
T^{\omega<m}_{uu} \approx \int \limits_{\mu<|\omega^{\prime}|<m}\dfrac{d\omega^{\prime}}{2\pi}\int \limits_{\mu<|\omega^{\prime \prime}|<m}\dfrac{d\omega^{\prime \prime}}{2\pi}\omega^{\prime \ 2} r_{g}e^{2\pi r_{g}\omega^{\prime}}|\Gamma(-2i\omega^{\prime}r_{g})|^{2}\delta(\omega^{\prime}-\omega^{\prime\prime}) \approx \nonumber\\
\approx \dfrac{1}{2\pi}\int \limits_{\mu<|\omega^{\prime}|<m}\dfrac{d\omega^{\prime}}{2\pi}\omega^{\prime \ 2} r_{g}e^{2\pi r_{g}\omega^{\prime}}\dfrac{2\pi}{2 \omega^{\prime} r_{g}\left(e^{2\pi \omega^{\prime}r_{g}}-e^{-2\pi \omega^{\prime}r_{g}}\right)} \approx \nonumber \\
\approx \dfrac{1}{4\pi}\int \limits_{\mu}^{m}d\omega  \ \omega+\int \limits_{\mu}^{m}\dfrac{d\omega}{2\pi}\dfrac{\omega}{e^{4\pi \omega r_{g}}-1} \approx \dfrac{m^{2}}{8\pi}+\int \limits_{0}^{m}\dfrac{d\omega}{2\pi}\dfrac{\omega}{e^{4\pi \omega r_{g}}-1}.
\end{gather}
Similarly, one can show that 
\begin{eqnarray}
\label{eq14!}
T_{vv} \approx \dfrac{m^{2}}{8\pi}+\int \limits_{0}^{m}\dfrac{d\omega}{2\pi}\dfrac{\omega}{e^{4\pi \omega r_{g}}-1},
\end{eqnarray}
where $\frac{m^{2}}{8\pi}$ exactly cancels unphysical part present in \eqref{eq:197f} and \eqref{eq:198fv}.
Finally, $T_{uu}$ component near the shell has the form
\begin{eqnarray}
\label{eq15!}
T_{uu}=T^{\omega<m}_{uu}+T^{\omega>m}_{uu} \approx \nonumber \\
\approx \dfrac{m^{2}}{8\pi}+\int \limits_{0}^{m}\dfrac{d\omega}{2\pi}\dfrac{\omega}{e^{4\pi \omega r_{g}}-1}-\int \limits_{0}^{+\infty}\dfrac{d\omega}{2\pi}\dfrac{\omega}{e^{4\pi \omega r_{g}}-1}-\dfrac{m^{2}}{8\pi}+\int \limits_{m}^{+\infty}\dfrac{d\omega}{2\pi}\dfrac{\omega}{e^{4\pi \omega r_{g}}-1} \approx 0.
\end{eqnarray}
Similarly,
\begin{eqnarray}
\label{eq16!}
T_{vv}\approx \dfrac{m^{2}}{8\pi}+\int \limits_{0}^{m}\dfrac{d\omega}{2\pi}\dfrac{\omega}{e^{4\pi \omega r_{g}}-1}-\int \limits_{0}^{+\infty}\dfrac{d\omega}{2\pi}\dfrac{\omega}{e^{4\pi \omega r_{g}}-1}-\dfrac{m^{2}}{8\pi} \approx -\int \limits_{m}^{+\infty}\dfrac{d\omega}{2\pi}\dfrac{\omega}{e^{4\pi \omega r_{g}}-1}
\end{eqnarray}
Finally, we get that 
\begin{eqnarray}
\label{eq17!}
 \left\{ \begin{matrix}
    T_{uu} \approx 0, \\
    T_{vv} \approx -\int \limits_{m}^{+\infty}\dfrac{d\omega}{2\pi}\dfrac{\omega}{e^{4\pi \omega r_{g}}-1}, \\
    T_{tr_{*}} \approx \int \limits_{m}^{+\infty}\dfrac{d\omega}{2\pi}\dfrac{\omega}{e^{4\pi \omega r_{g}}-1}
    \end{matrix}\right. \qquad \qquad r\rightarrow r_{g}.
\end{eqnarray}


The calculation of $T_{uv}$ near the shell is very similar to the one presented in Appendix \ref{Calculations}. Using the calculations from the appendix, we can find contribution from the modes with $\omega>m$: 
\begin{gather}
    \label{d30f}
    \dfrac{1}{2}\bigg\langle\phi(x^{+})\phi(x^{-})+\phi(x^{-})\phi(x^{+})\bigg\rangle_{\omega>m} \approx \nonumber\\
    \approx \int\limits_{m}^{+\infty}\dfrac{d\omega}{2\pi}\dfrac{\cos{\left(\omega[u^{+}-u^{-}]\right)}}{2k(r_{*})}+\int\limits_{m}^{+\infty}\dfrac{d\omega}{2\pi}\dfrac{\cos{\left(\omega[v^{+}-v^{-}]\right)}}{2k(r_{*})}+\int \limits_{m}^{+\infty}\dfrac{d\omega }{2\pi}\dfrac{\omega^{-1}}{(e^{4\pi r_{g}\omega }-1)}.
\end{gather}
The main difference is that in the vicinity of the shell we also need to take into the account the harmonics \eqref{d27f} from the region $\omega<m$. Again, as in appendix \ref{SEThorizon}, after integrating over the variable $\omega$, we calculate the contribution of $\delta(\omega^{\prime}-\omega^{\prime \prime})$ in \eqref{eq9!} only, since other contributions lead to terms that decay as powers of $1/u$. Repeating steps from the previous appendixes with the harmonics \eqref{d27f} we obtain
\begin{eqnarray}
\label{eq101!}
\dfrac{1}{2}\bigg\langle\phi(x^{+})\phi(x^{-})+\phi(x^{-})\phi(x^{+})\bigg\rangle_{\omega<m}
\approx \nonumber\\
\approx \lim_{\mu 
    \to 0}\int \limits_{\mu<|\omega|<m}\dfrac{d\omega}{2\pi}\dfrac{1}{4\omega}\dfrac{1}{1-e^{-4\pi r_{g} \omega}}\left[e^{2i\omega r_{*}+i\pi}+e^{i\omega[u^{+}-u^{-}]}+e^{i\omega[v^{+}-v^{-}]}+e^{-2i\omega r_{*}-i\pi}\right]
\end{eqnarray}
This expression for $T_{uv}$ should be compared with the equation $(4.12)$ in \cite{Akhmedov:2020ryq}. Following this paper we change $\int \limits_{\mu<|\omega|<m} \rightarrow \int \limits_{-m}^{m}$. It is again worth to mention that we can not find the modes with $0<\omega<\mu$ because we cannot solve equation \eqref{eq6}. But as we show below we obtain the right Hadamard term in $T_{uv}$ which justifies our approximation. 
To calculate integrals, present in \eqref{eq101!}, we use regularization from \cite{Akhmedov:2020ryq} and make the change $\omega \rightarrow \omega+is$ after which we use Jordan's lemma to calculate integrals. The results in the limit $mr_{g} \gg 1$ and $mr_{*} \rightarrow -\infty$ has the form
\begin{gather}
\label{tuveq10}
\lim_{s \to 0}\int \limits_{-m}^{m}\dfrac{d\omega}{2\pi}\dfrac{1}{4(\omega+is)}\dfrac{1}{1-e^{-4\pi r_{g} (\omega+is)}}\left[e^{2i\omega r_{*}+i\pi}+e^{i\omega[u^{+}-u^{-}]}+e^{i\omega[v^{+}-v^{-}]}+e^{-2i\omega r_{*}-i\pi}\right]+\text{c.c} = -\dfrac{1}{4\pi}\dfrac{r_{*}}{r_{g}}
\end{gather}
Using that when $r_{*}\rightarrow -\infty$ the geodesic distance between points $x^{+}$ and $x^{-}$ has the form
\begin{eqnarray}
\label{tuveq11}
\sigma^{2}=\left(1-\dfrac{r_{g}}{r(r_{*})}\right)[u^{+}-u^{-}][v^{+}-v^{-}] \approx e^{\frac{r_{*}}{r_{g}}}[u^{+}-u^{-}][v^{+}-v^{-}],
\end{eqnarray}
and taking the logarithm of \eqref{tuveq11} we get that
\begin{eqnarray}
\label{tuveq12}
\log{\sigma^{2}} \approx \dfrac{r_{*}}{r_{g}}+\log{[u^{+}-u^{-}][v^{+}-v^{-}]}.
\end{eqnarray}
Finally, using the asymptotic form of cosine integral
\begin{eqnarray}
\label{tuveq13}
\lim_{z \to 0}\int\limits_{z}^{+\infty}dx\dfrac{cos{x}}{x}=-\lim_{z \to 0}Ci(z) \approx -\gamma -\log{z},
\end{eqnarray}
where $\gamma$ is the Euler-Mascheroni constant, and using that the integral  $\int \limits_{m}^{+\infty}\dfrac{d\omega}{2\pi}\dfrac{1}{\omega}\dfrac{1}{e^{4\pi r_{g}\omega}-1}$ is negligibly small in the limit $mr_{g}\gg 1$, we get that
\begin{gather}
\label{tuveq14}
T_{uv} \approx \dfrac{m^{2}}{8\pi}g_{uv}\bigg[-\log{\sigma^{2}}+\log{\left([u^{+}-u^{-}][v^{+}-v^{-}]\right)}-\log{\left(m^{2}[u^{+}-u^{-}][v^{+}-v^{-}]\right)}-2\gamma\bigg]\approx \nonumber \\
=-\dfrac{m^{2}}{4\pi}g_{uv}\bigg[\log{\left(m\sigma\right)}+\gamma\bigg]
\end{gather}
Once again, we want to emphasize the point made in \cite{Akhmedov:2020ryq}: the phase $\pi$ (in the notations of \cite{Akhmedov:2020ryq} it is $2\delta_{0}$) as $\omega \rightarrow 0$ at \eqref{tuveq10} is very important: only for such a phase it is possible to get the correct singularity for $T_{uv}$. Indeed, if we formally change $\pi\rightarrow 2\delta_{\omega}$ in \eqref{tuveq10}, we obtain:
\begin{gather}
\label{tuveq15}
\lim_{s \to 0}\int \limits_{-m}^{m}\dfrac{d\omega}{2\pi}\dfrac{1}{4(\omega+is)}\dfrac{1}{1-e^{-4\pi r_{g} (\omega+is)}}\left[e^{2i\omega r_{*}+i2\delta_{\omega}}+e^{i\omega[u^{+}-u^{-}]}+e^{i\omega[v^{+}-v^{-}]}+e^{-2i\omega r_{*}-i2\delta_{\omega}}\right]+\text{c.c} \approx \nonumber\\
\approx \dfrac{1}{4}\sin{\left(2\delta_{0}\right)}+\dfrac{r_{*}}{4\pi r_{g}}\cos{\left(2\delta_{0}\right)},
\end{gather}
which gives $-\dfrac{1}{4\pi}\log{(m\sigma)}$ only for $2\delta_{0}=\pi$. Also, as was mentioned above, we cannot find the exact form of modes with $0<\omega<\mu$ which play a crucial role since they contain $2\delta_{0}$. So, in some sense after taking the limit $\mu \rightarrow 0$ we solve an inverse problem: we define such $2\delta_{0}$ which gives standard UV singularity for $T_{uv}$.

\end{document}